\documentclass[aps,prl,superscriptaddress,reprint]{revtex4-1}

\usepackage{graphicx}
\usepackage{epstopdf,epsfig,subfigure}
\usepackage{amsmath,amssymb,latexsym}
\usepackage{empheq}
\usepackage{url}
\usepackage{bm} 
\usepackage{lineno,color}

\begin{document}


\title{A critical point for bifurcation cascades and featureless turbulence}

\author{Jacopo Canton}
\email[]{jcanton@ethz.ch}
\affiliation{%
    Computational Science and Engineering Laboratory,
	ETH Zurich,
	CH-8092 Zurich, Switzerland
}
\affiliation{%
    Linn\'e FLOW Centre
	KTH Mechanics,
	Royal Institute of Technology,
	SE-100 44 Stockholm, Sweden
}
\author{Enrico Rinaldi}
\email[]{erinaldi@mech.kth.se}
\affiliation{%
    Linn\'e FLOW Centre
	KTH Mechanics,
	Royal Institute of Technology,
	SE-100 44 Stockholm, Sweden
}
\author{Ramis \"Orl\"u}
\email[]{ramis@mech.kth.se}
\affiliation{%
	Linn\'e FLOW Centre
	KTH Mechanics,
	Royal Institute of Technology,
	SE-100 44 Stockholm, Sweden
}
\author{Philipp Schlatter}
\email[]{pschlatt@mech.kth.se}
\affiliation{%
	Linn\'e FLOW Centre
	KTH Mechanics,
	Royal Institute of Technology,
	SE-100 44 Stockholm, Sweden
}

\date{\today}

\begin{abstract}
In this Letter we show that a bifurcation cascade and fully sustained turbulence can share the phase space of a fluid flow system, resulting in the presence of competing stable attractors.
We analyse the toroidal pipe flow, which undergoes subcritical transition to turbulence at low pipe curvatures (pipe-to-torus diameter ratio) and supercritical transition at high curvatures, as was previously documented.
We unveil an additional step in the bifurcation cascade and provide evidence that, in a narrow range of intermediate curvatures, its dynamics competes with that of sustained turbulence emerging through subcritical transition mechanisms.
\end{abstract}


\maketitle 


The origin of turbulence is one of the outstanding problems in classical physics and dynamical systems theory.
More than 130 years after the seminal experiments by Osborne Reynolds, it is still a matter of debate~\cite{Reynolds:1883cq,2016_Pomeau}.
Not only is it of scientific and engineering interest for the fluids community~\cite{Barkley:2019fv}, but it also serves as a proving ground for the study of open complex dynamical systems and chaos theory.
Many systems appearing in Nature present multiple solutions and exhibit complex routes to chaos, examples include the magnetic cycles of stellar dynamos~\cite{Knobloch:df}, large-scale oscillations and bursts in tokamak plasmas~\cite{Beyer:2005cp}, as well as a number of physico-chemical systems~\cite{Moss:1994kl}.
A flow can naturally transition to turbulence as a consequence of linear instabilities (supercritical transition).
However, turbulence can also appear in linearly stable flows subjected to sufficiently strong perturbations (subcritical, or by-pass, transition).
These two scenarios do not typically coexist in hydrodynamical systems and whether one observes supercritical or subcritical transition depends on the characteristics of the dynamical system or the governing parameters~\cite{Reshotko:2001he,Beneitez:2019gx}; evidence of their interaction is scarce~\cite{Meseguer:2009jl,Crowley2019}.

The flow between two rotating cylinders---Taylor--Couette flow---is a classic example of transition initiated by linear instabilities (see Ref.~\cite{1981_Eckmann} and references therein).
When a governing parameter of the system---typically the Reynolds number $Re$ in fluid dynamics---is increased past a critical value, the system undergoes a supercritical Hopf bifurcation and the stable state of the flow changes from a steady fixed point to a periodic limit cycle.
A further increase of the governing parameter causes a secondary Hopf (Neimark--Sacker) bifurcation%
~\footnote{A bifurcation occurring when the fixed point of a discrete-time map changes stability via a pair of complex-conjugate eigenvalues (multipliers) with unit modulus
\cite{Neimark1959,Sacker1964,Kuznetsov}.}%
, which renders  the limit cycle unstable and introduces a new frequency in the flow, increasing its spatio--temporal complexity.
The system can reach a chaotic state following a third bifurcation.
Taylor--Couette flow, in certain configurations, actually follows a Ruelle--Takens scenario
\footnote{When a system undergoes three Hopf-type bifurcations, starting from a stationary solution as a parameter is varied, then it is likely that the system possesses a strange attractor with sensitivity to initial conditions after the third bifurcation.
The power spectrum of such a system will exhibit one, then two, and possibly three independent basic frequencies.
When the third frequency is about to appear, simultaneously some broad-band noise will appear if there is a strange attractor~\cite{Ruelle:1971jk,Newhouse:1978fl,1981_Eckmann}.} as was established experimentally in Ref.~\cite{Gollub:1975gy}.

In linearly stable flows, on the other hand, subcritical bifurcations may occur and transition from a state to another is triggered by the introduction of finite-amplitude perturbations.
This is a predominantly nonlinear process that relies on the existence of a non-trivial set other than the laminar state~\cite{1995_Waleffe}.
Relaminarising and transitional trajectories are separated in phase space by a manifold known as the ``edge of chaos''~\cite{2006_SkufcaEtAl,2007_SchneiderEtAl,2012_deLozarEtAl}.
The trajectories that make their way to the turbulent attractor are generally organised around a set of unstable solutions, e.g., travelling waves or relative periodic orbits, at least at low Reynolds numbers~\cite{2012_KawaharaEtAl,2017_BudanurEtAl}.
Flows in channels and pipes are examples of this scenario, and are dominated by large-scale spatial and temporal intermittency at the onset of turbulence.
Only recently theoretical models have been shown to capture the complex physics of transition and match laboratory measurements~\cite{Barkley2015,Shih:2015dl,Barkley2016a}.
Notably, there is an increasing body of experimental and numerical evidence suggesting an analogy between subcritical transition and non-equilibrium phase transition of the directed percolation type~\cite{Shih:2015dl,2016_LemoultEtAl,2017_ChantryEtAl}.
The accuracy of this analogy is still a matter of debate~\cite{Barkley2016a,Mukund2018}, but it is clear that flow structures encountered in subcritical and supercritical transition scenarios are fundamentally different in nature, and their description is rooted in distinct theoretical grounds.

Only a few flow cases present both transition scenarios.
Taylor--Couette flow undergoes a sequence of supercritical bifurcations when the cylinders are co-rotating, while subcritical transition is observed when the cylinders are counter-rotating~\cite{Coles1965,1981_Eckmann,Andereck1986}.
A similar behaviour is observed in rotating plane Couette flow~\cite{Tsukahara:2010jh}.
In both cases, however, the two transition scenarios were not known to interact with each other until the present study and one contemporary to it~\cite{Crowley2019}.
In spatially developing boundary layers recent work suggests that at high Reynolds numbers the edge of chaos can effectively be interpreted as a manifold dividing classical supercritical and subcritical bypass-type transition \cite{Beneitez:2019gx}.

\begin{figure}
	\centering
	\includegraphics[width=\columnwidth]{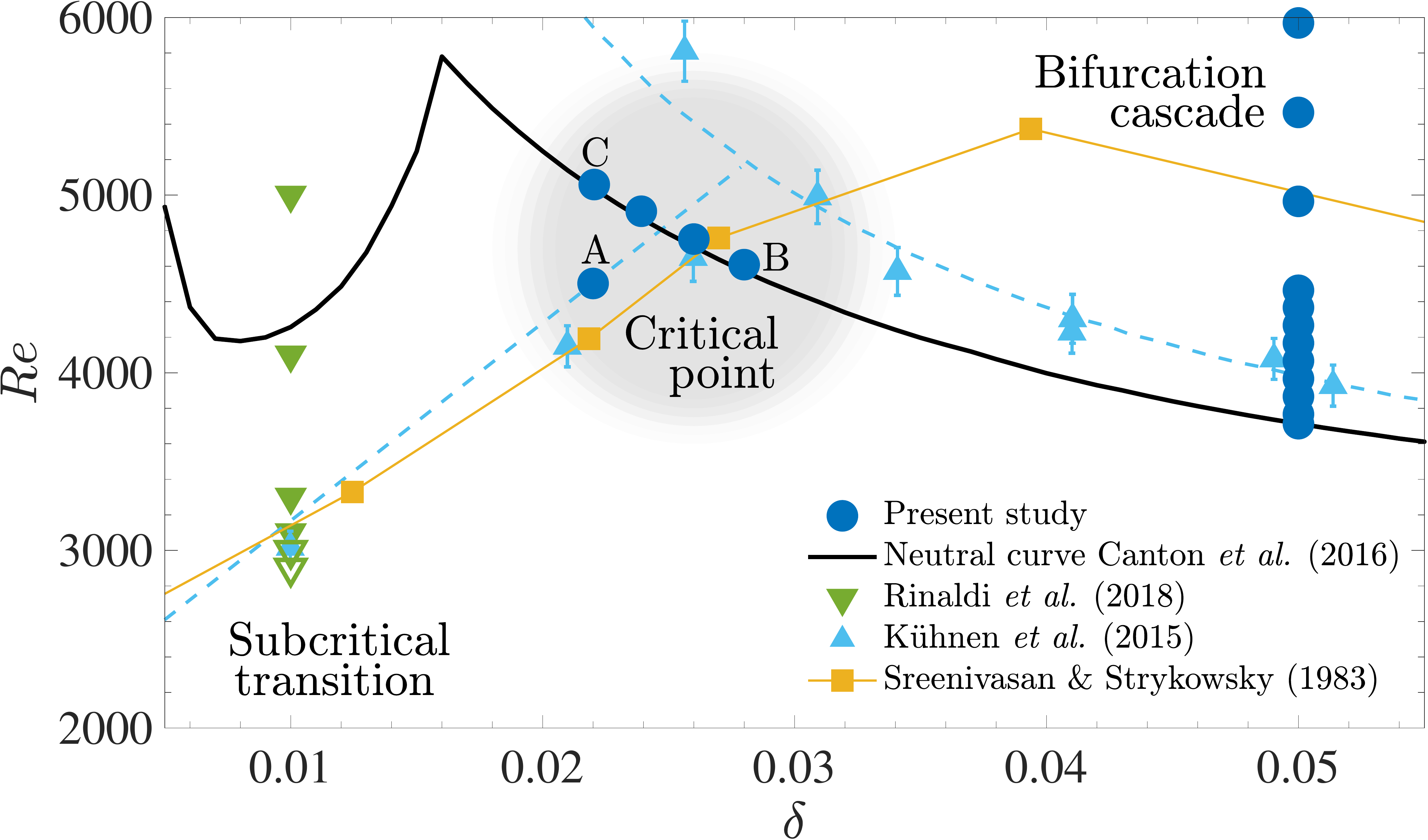}
	\caption{%
		Portion of the $\delta-Re$ parameter space of the flow in a toroidal pipe. Experimental and numerical data from the literature are reported as well as the location of the present computations.
        The gray shaded area indicates the approximate boundaries of the critical point, where subcritical transition and a bifurcation cascade coexist.
		Filled downward-pointing triangles indicate spatially expanding turbulence, while empty triangles relaminarization.
		The data from Sreenivasan \& Strykowsky (Ref.~\cite{Sreenivasan1983}) is the curve they refer to as the ``conservative lower critical limit''.
		The data from K\"uhnen et al.\ (Ref.~\cite{Kuhnen2015}) denote $50\%$ intermittency at low curvatures and the appearance of the supercritical travelling wave above the critical point.
    }\label{fig:neutralCritical}
\end{figure}

The flow in a bent pipe is a rare case in which the nature of transition is altered without such a clear separation.
The change takes place as the pipe-to-torus diameter ratio $\delta = D/D_T$ is gradually increased~\cite{Kuhnen2015} and will be the focus of this Letter.
Figure~\ref{fig:neutralCritical} sets the stage by presenting an overview of this flow case in the ($\delta$, $Re$) parameter space.
Transition to turbulence is subcritical at low curvatures and qualitatively similar to the one in straight pipes~\cite{Rinaldi:2019ab,Sreenivasan1983,Kuhnen2015}.
For larger $\delta$, instead, transition is initiated by a supercritical Hopf bifurcation~\cite{Sreenivasan1983,Kuhnen2014,Kuhnen2015,Canton:2016in} and all elements point towards a bifurcation cascade~\cite{Kuhnen2014}.
The dynamics of the flow at intermediate curvatures, however, remains largely unexplored, leaving unanswered the question of whether the two transition scenarios interact, and if so how.
In this Letter we address this question and show that characteristic structures of a bifurcation cascade and a subcritical transition scenario can coexist at a fixed combination of all governing flow parameters.

For the same combination of $\delta$ and $Re$, we find two competing attractors with complementary basins, namely sustained turbulence and a stable travelling wave (relative equilibrium) originated by the supercritical Hopf bifurcation predicted by linear theory.
This is in contrast with what can be observed, e.g., in a linearly unstable channel flow, where only one asymptotic state exists regardless of whether transition is caused by Tollmien--Schlichting waves or by by-pass mechanisms%
~\footnote{Above the critical Reynolds number $Re \approx 5772$ channel flow becomes linearly unstable and the critical mode of instability is called Tollmien--Schlichting wave. Transition to turbulence may occur through the breakdown of such waves for carefully chosen initial conditions. However, in the presence of other disturbances of finite amplitude, the critical mode is by-passed by the nucleation and expansion of localised turbulent patches~\cite{SchmidHenningson}}.
Bent pipes are not the first flow case for which two competing stable---but not necessarily steady---solutions are documented, see, e.g.\ Taylor--Couette flow~\cite{Coles1965} or the flow through a sinuous stenosis~\cite{Samuelsson:2015cq}.
However, the competing solutions in Taylor--Couette appear to be either all of supercritical type or reached through a hysteresis cycle~\cite{Crowley2019}, while those in a stenotic flow belong to two branches originated by the same saddle--node bifurcation, and do not bear the imprint of the two fundamentally different transition scenarios.
We perform direct numerical simulations (DNS) of the flow in a toroidal pipe%
~\footnote{The incompressible Navier--Stokes equations are solved using the open-source code \textsc{Nek5000} \cite{nek5000}. The numerical setup was previously validated for straight and bent pipes \cite{ElKhoury:2013kj,Noorani2013}.}
and articulate our analysis in two steps. 

First, we shed light on the bifurcation cascade at large curvatures, and provide evidence for a secondary Hopf bifurcation.  The occurrence of a second bifurcation was ``conjectured'' in Ref.~\cite{Kuhnen2015} but  ``could not be pinpointed'' experimentally (quotations from Refs.~\cite{Kuhnen2014,Kuhnen2015}).
Second, we move our attention to the region of the parameter space where the neutral curve of the flow intersects the threshold for subcritical transition%
~\footnote{Defined as 50\% intermittency in Ref.~\cite{Kuhnen2015}}, hereafter referred to as \textit{critical point}, see Fig.~\ref{fig:neutralCritical}.


\begin{figure}
	\centering
	\includegraphics[width=\columnwidth]{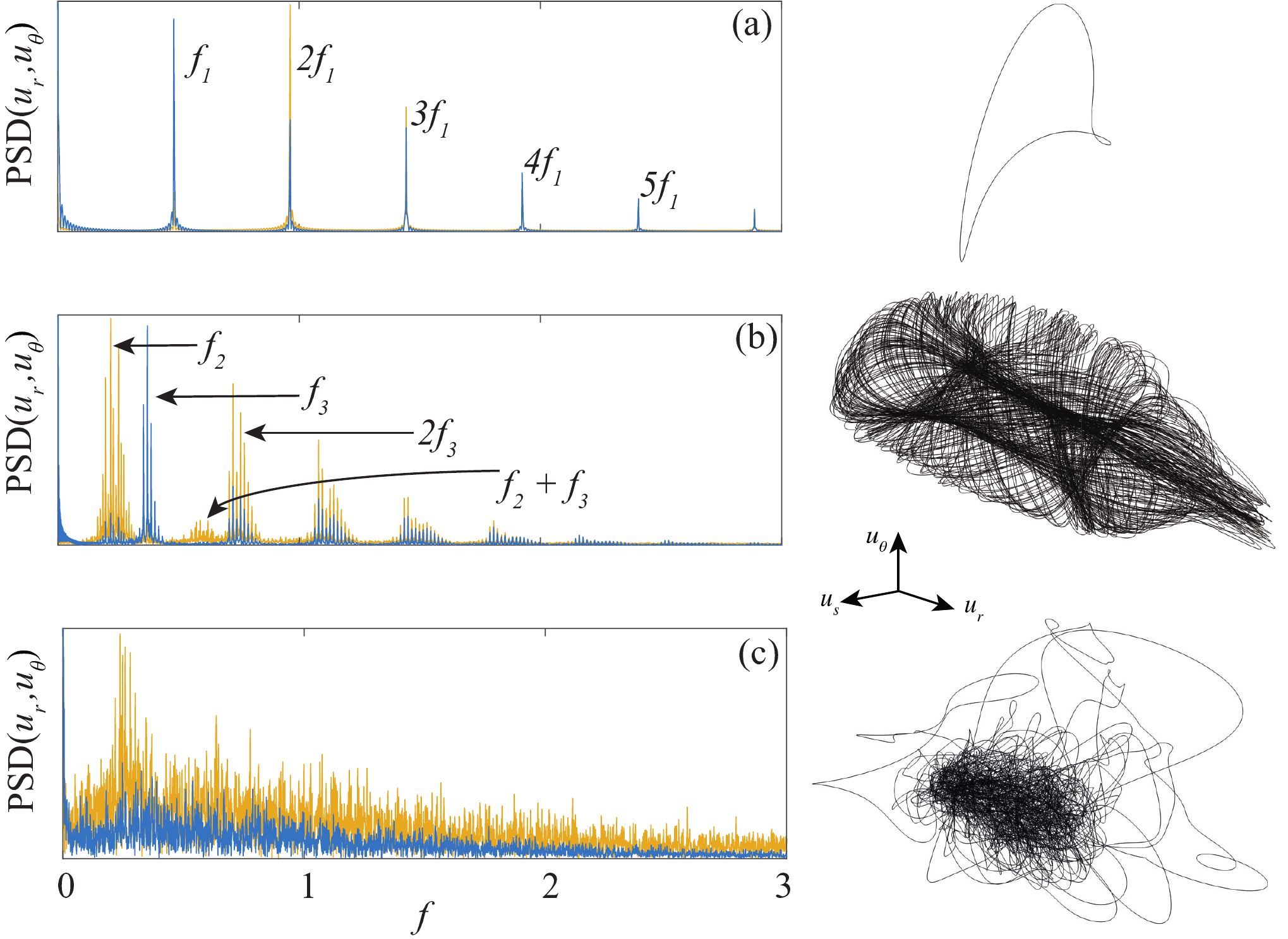}
	\caption{%
		Power spectral densities (PSD) and corresponding phase space representations for
		point velocity measurements at
		$\delta=0.05$ and $Re=4000$ (a), $Re=4500$ (b), and $Re=6000$ (c).
		Yellow lines represent the PSD of radial velocity $u_r$, while blue lines that of the azimuthal component $u_\theta$.
	}\label{fig:supercritical_spectra}
\end{figure}

\begin{figure*}
    \centering
    \includegraphics[width=\textwidth]{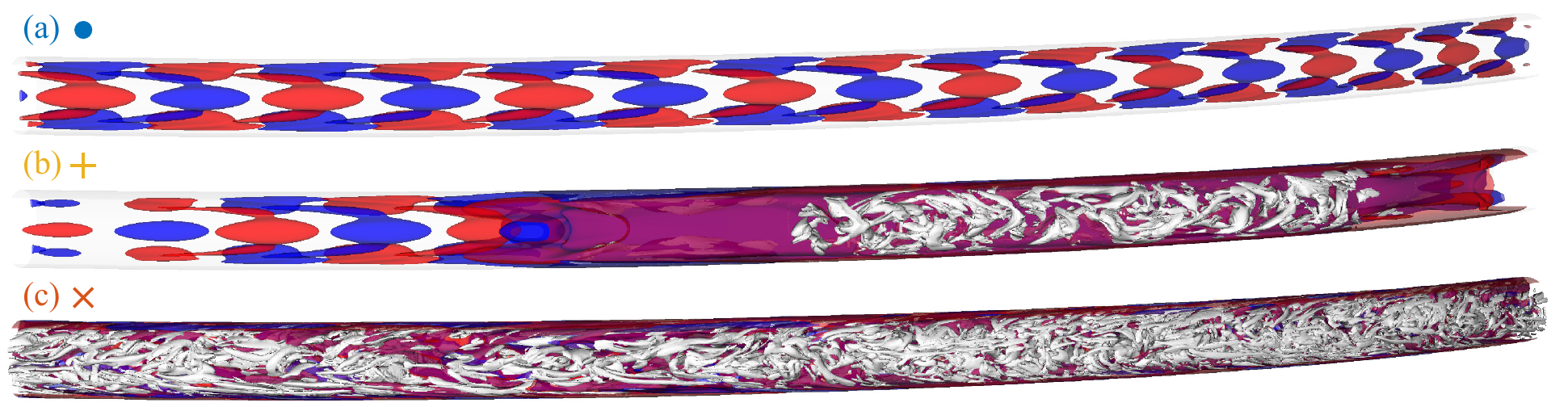}
    \caption{%
	    Snapshots of the flow for $\delta=0.022$, $Re=5050$ along the three trajectories in Fig.~\ref{fig:critical_phasespace}.
	    (a) corresponds to the (blue) travelling wave, (b) to the intermediate (orange) trajectory that returns to the travelling wave, while (c) is along the (red) trajectory that converges to sustained turbulence.%
    }
\label{fig:snapshots}
\end{figure*}

\begin{figure}
    \centering
    \includegraphics[width=\columnwidth]{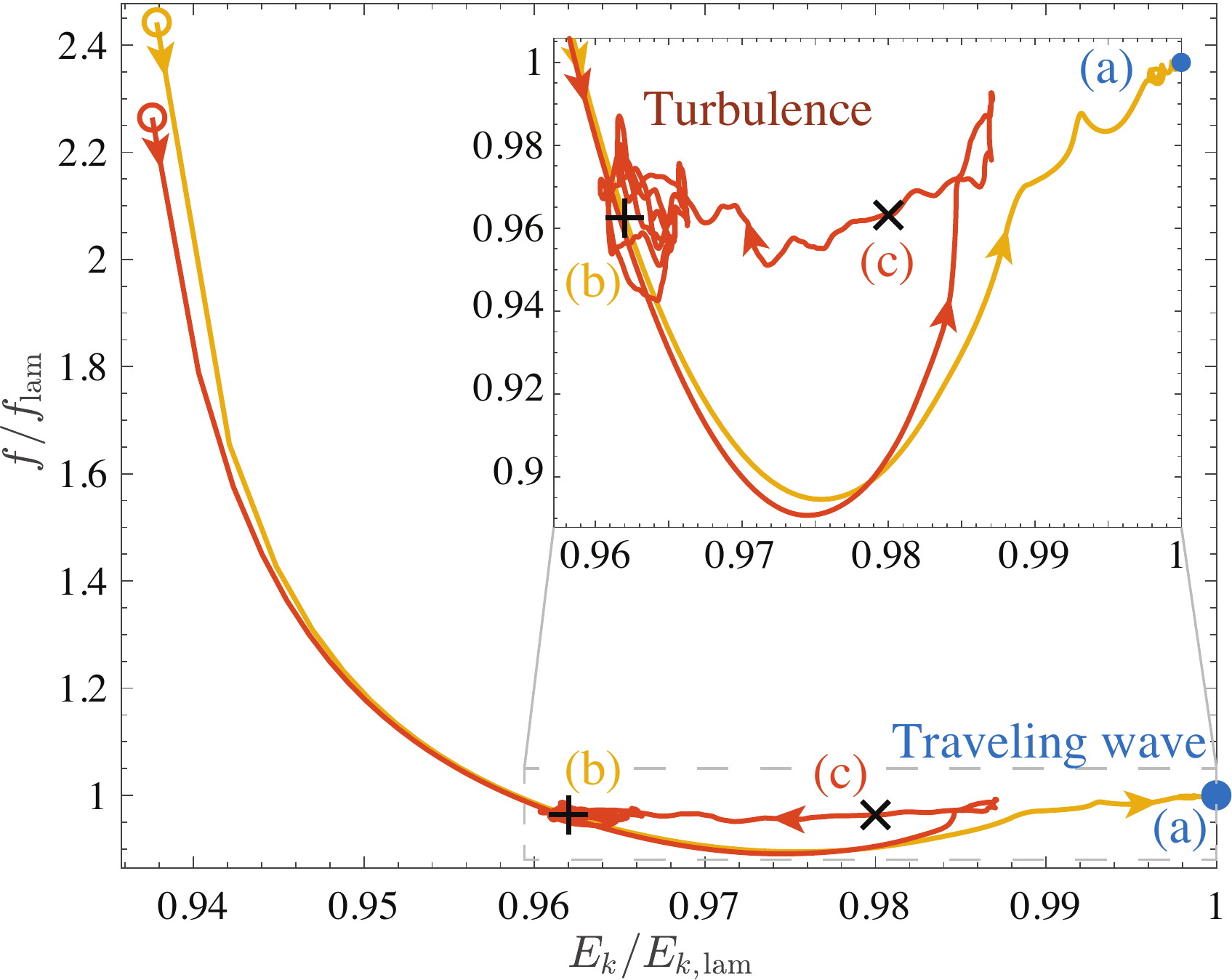}
    \caption{%
    	Trajectories in the kinetic energy--friction factor phase space at the critical point, $\delta=0.22$, $Re=5050$.
    	Arrowheads indicate the direction of time, and $E_k$ and $f$ are normalized by their respective values for the (unstable) laminar flow.
    	The filled circle, plus and cross markers indicate the location of the snapshots in Fig.~\ref{fig:snapshots}.
    	It is expected that for turbulent flow both $E_k$ and $f$ are lower than for the laminar flow, as this curvature range is subject to sublaminar drag~\citep{Noorani2015}.
    	Being just above the neutral curve, $Re=5013$ at this curvature, the travelling wave is still very close ($<1\%$) to the unstable laminar flow.
    }\label{fig:critical_phasespace}
\end{figure}
   
Regarding the bifurcation cascade, we focus on curvature $\delta=0.05$, which is far enough from the critical point to guarantee no influence of the subcritical transition scenario.
By means of a modal stability analysis we verified that the steady flow becomes linearly unstable at $Re=3713$ and nonlinear simulations up to $Re=4000$ confirmed the nature of the supercritical Hopf bifurcation~\cite{Canton:2016in}.
Figure~\ref{fig:supercritical_spectra}(a) depicts the state of the system at this Reynolds number: the flow is constituted by a travelling wave generated by the first supercritical Hopf bifurcation and all trajectories in the phase space eventually converge to the corresponding relative equilibrium.
The power spectral density (PSD) of the velocity signal at a fixed point presents an isolated peak at frequency $f_1\approx0.48$, followed by its harmonics (made dimensioneless with bulk velocity and pipe diameter, i.e.\ $f=f^*D/U$).
On the other hand, a PSD analysis of integral quantities, such as the kinetic energy $E_k$ and friction factor $f$, reveals no frequencies (except zero frequency).
This is expected for systems with continuous symmetries, as the true dimension of the system is revealed in a symmetry-reduced phase space only---such as that constituted by $E_k$ and $f$---see e.g.\ Ref.~\cite{Willis:2013bu}.
Upon increasing the Reynolds number well beyond the critical one, the system undergoes a secondary Hopf bifurcation to what appears as a quasi-periodic state in the velocity state space, but is actually a relative periodic orbit in a symmetry-reduced state space.
The attractor appears as a $T^2$ torus in the velocity space, depicted in figure~\ref{fig:supercritical_spectra}(b) for $Re=4500$, and the PSD of the point velocity signal presents two incommensurable frequencies, $f_2\approx0.22$ and $f_3\approx0.37$, with their linear combinations and higher harmonics.
Once again, the PSD of integral quantities shows one less frequency, indicating that the flow is on a relative periodic orbit.
As the Reynolds number is increased further the flow eventually becomes turbulent in the whole pipe.
The trajectory in the corresponding phase space is chaotic and the PSD of both point- and integral quantities is characterised by a broadband spectrum, see Fig.~\ref{fig:supercritical_spectra}(c).
Further bifurcations could not be accurately pinpointed with additional simulations between $Re=4500$ and $6000$, but only the appearance of broadband noise and eventually chaos.
If this system were to follow a Ruelle--Takens route, either one of two scenarios could be observed between $Re=4500$ and $6000$:
a Neimark--Sacker bifurcation would lead the system to a $T^2$ torus---in the $E_k-f$ phase space---followed either by a direct transition to chaos, or by a fourth bifurcation to a $T^3$ torus before the appearance of chaos.
However, three-frequency tori are rare in hydrodynamical systems, albeit not impossible~\cite{Oteski:2015ie}.
We conclude that, for curvatures sufficiently larger than the ones at which subcritical transition occurs, from $\delta\approx0.025$ up to unity, the flow indeed undergoes a bifurcation cascade that leads it to chaos, similarly to what is observed for Taylor--Couette~\cite{Coles1965}.


We now turn our attention to the critical point, i.e.\ the area within $0.02\lesssim\delta\lesssim0.03$ and $4000\lesssim Re \lesssim 5000$.
These boundaries are approximate, as is the grey shaded area in Fig.~\ref{fig:neutralCritical}.
In this region of the parameter space we perform DNS in domains of length $L_s = 4$--$8\,\lambda_s$---corresponding to approximately $10$--$20D$---where $\lambda_s$ is the wavelength of the travelling wave predicted by the linear theory at each curvature%
~\footnote{These domains are long enough for accommodating sustained turbulence as well as over 4 periods of the fundamental travelling wave.  Note that turbulence is not intermittent in this region of the parameter space~\cite{Sreenivasan1983}.
When not simulating the whole torus, the domain length is restricted to integer multiples of the fundamental wavelength such that any dynamics present in these domains would also appear in the full toroidal geometry.
This restriction does not introduce additional feedback effects, nor does it alter the nature of transition as was previously observed in Refs.~\cite{Noorani2013,Noorani2015,Canton:2016in,Rinaldi:2019ab}.%
}.
As a first step we verify that both sub- and supercritical behaviours can still be isolated.
In order to check for subcritical behaviour we perform DNS for $\delta=0.022$ and $Re=4500$ (point A in Fig.~\ref{fig:neutralCritical}).
Each simulation is initialised with a single turbulent puff, selected from the ones computed in Ref.~\cite{Rinaldi:2019ab}, which expands---becoming a slug at this $Re$---and fills the whole computational domain with turbulent flow, in agreement with the findings in Refs.~\cite{Sreenivasan1983,Kuhnen2015,Rinaldi:2019ab}.
At this Reynolds number, in fact, we are above the intermittency range where turbulent puffs are observed~\cite{Sreenivasan1983,Kuhnen2015,Rinaldi:2019ab}.  An appropriate perturbation is sufficient to initiate the by-pass mechanism that renders the flow homogeneously turbulent.
The second test, to check for supercritical behaviour, is at $\delta=0.028$ and $Re=4600$ (point B in Fig.~\ref{fig:neutralCritical}).
This point is located below the subcritical transition threshold and just above the neutral curve---which for this curvature marks the first Hopf bifurcation at $Re=4570$.
The flow is initialised with a parabolic streamwise velocity profile perturbed with either random noise or the unstable eigenmode, as was done in Ref.~\cite{Canton:2016in}---in both cases increasing the energy of the flow field by approximately 1\%.
These simulations always converge to the nonlinear travelling wave created by the supercritical Hopf bifurcation, confirming the supercritical transition route previously discussed for $\delta=0.05$.

Having established both behaviours---subcritical in point A and supercritical in B---close to criticality, we explore the region above the neutral curve between points B and C.
We investigate three pairs of $(\delta, Re)$, i.e.\ $(0.026, 4750)$, $(0.024, 4900)$, and $(0.022, 5050)$, the latter corresponding to point C in Fig.~\ref{fig:neutralCritical}, which is analysed in detail in the following and in Figs.~\ref{fig:snapshots} and \ref{fig:critical_phasespace}.

Unlike in other regions of the parameter space, for $(\delta, Re) = (0.022, 5050)$ the flow shows sensitivity to initial conditions, which dictate its asymptotic state.
For these critical values of $\delta$ and $Re$ an initial condition consisting of a randomly perturbed parabolic velocity profile slowly converges to a stable nonlinear travelling wave, as predicted by the modal analysis and as observed for higher curvatures.
The flow trajectory projected on the energy--friction phase space collapses to a single point, illustrated in blue in Fig.~\ref{fig:critical_phasespace}.
On the other hand, if the laminar flow is perturbed with a \textit{sufficiently energetic} localised disturbance---a puff in our case---the disturbance grows and invades the whole length of the pipe, turning the flow into a persistently turbulent state~%
\footnote{Experiments in Refs.~\cite{Sreenivasan1983,Kuhnen2015} indicate persistent turbulence for these values of $\delta$ and $Re$.}
through a subcritical-like transition process (red line in Fig.~\ref{fig:critical_phasespace}).
For this particular combination of $\delta$ and $Re$, \textit{sufficiently energetic} means that the kinetic energy of the puff accounts for more than 4.8\% of the kinetic energy of the flow field.
If the puff has lower initial energy, it will only transiently grow in size before eventually disappearing (orange line in Fig.~\ref{fig:critical_phasespace}).
In this case the flow state visits the neighbourhood of the turbulent attractor in phase space, but eventually converges to the travelling wave.
Snapshots of the flow taken from these three different trajectories are reported in Fig.~\ref{fig:snapshots} which visually illustrates the temporary coexistence of travelling wave and puff.

\begin{figure}
    \centering
    \includegraphics[width=\columnwidth]{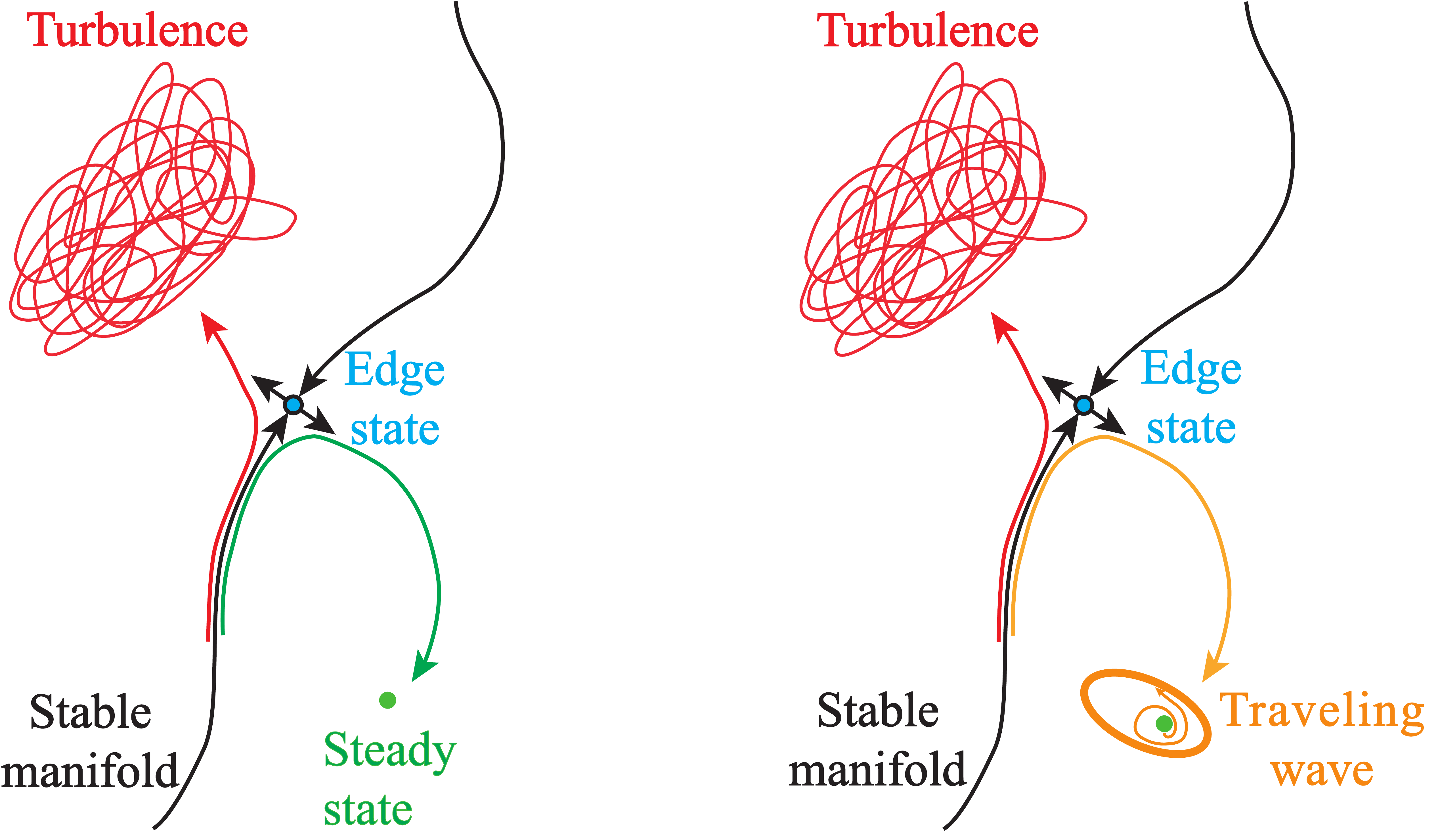}
    \caption{%
    	Sketch of the phase space at criticality.
    	The left pane illustrates the behaviour of the system before the bifurcation: the steady state is stable and turbulence is reached via subcritical transition---point A in Fig.~\ref{fig:neutralCritical}.
    	The right pane shows the state of the system after the supercritical Hopf bifurcation: the steady state is now unstable and the first step of a bifurcation cascade has appeared in the form of a travelling wave---point C in Fig.~\ref{fig:neutralCritical}, illustrated in detail in Figs.~\ref{fig:snapshots} and \ref{fig:critical_phasespace}.
    	}\label{fig:sketch}
\end{figure}

The flow behaviour discussed in the present work clearly shows that a narrow region of the parameter space $(\delta, Re)$ presents two attractors, evidence of two completely different system dynamics, which compete and have complementary and finite basins of attraction.
The two attractors represent sustained turbulence and the travelling wave originated by a supercritical Hopf bifurcation. 
In analogy with the generally accepted picture of the phase space for subcritical flows, we sketch a modified phase space in Fig.~\ref{fig:sketch} where the laminar, steady-state attractor is replaced by the travelling wave and a saddle state acts as a mediator between flow trajectories.
The two attractors documented in this Letter embody two entirely different transition scenarios, which are shown to coexist in a fluid system, and lead the flow to two diametrically opposed unsteady asymptotic states.

\begin{acknowledgments}
The authors would like to acknowledge Y.\ Duguet, D.\ Barkley  and J.\ K\"uhnen for the helpful comments and discussions when preparing this manuscript.
Financial support by the Swedish Research Council (VR) and the Knut and Alice
Wallenberg Foundation (KAW) via the Wallenberg Academy Fellow programme is gratefully
acknowledged.
Computer time was provided by the Swedish National Infrastructure for Computing (SNIC).
\end{acknowledgments}


%
\end{document}